\documentclass[%
reprint,
showpacs,%
showkeys,
 amsmath,amssymb,
 aps,
pre,
longbibliography, floatfix,
]{revtex4-1}

\usepackage[english]{babel}

\usepackage{graphicx}

\newcommand{\onefig}[1]{\centering{\includegraphics[width=0.95\columnwidth]{#1}}}

\renewcommand{\Re}{\mathop{\mathrm{Re}}}
\renewcommand{\Im}{\mathop{\mathrm{Im}}}
\newcommand{\partialt}[1]{\partial_t #1}

\newcommand{\partialxx}[1]{\partial^2_x #1}

\newcommand{\myii}{\mathrm{i}}

\newcommand{\fdosv}{\nu_\tau^{(i)}}
\newcommand{\lyapdim}{D_\text{L}}
\newcommand{\oslspace}{\mathcal{S}}

\begin{document}

\title{Strict and fussy mode splitting in the tangent space of the Ginzburg-Landau equation}

\author{Pavel V. Kuptsov}\email[Corresponding author. Electronic address:]{p.kuptsov@rambler.ru}%
\altaffiliation[Permanent address:]{Department of Informatics, Saratov State Law Academy,
Volskaya 1, Saratov 410056, Russia}%

\author{Ulrich Parlitz}%
\affiliation{Drittes Physikalisches Institut,
Georg--August--Universi\"at G\"ottingen, Friedrich--Hund--Platz 1,
37073 G\"{o}ttingen, Germany}



\date{\today}

\begin{abstract}
In the tangent space of some spatially extended dissipative
systems one can observe ``physical'' modes which are highly
involved in the dynamics and are decoupled from the remaining set
of hyperbolically ``isolated'' degrees of freedom representing
strongly decaying perturbations. This mode splitting is studied
for the Ginzburg--Landau equation at different strength of the
spatial coupling. We observe that isolated modes coincide with
eigenmodes of the homogeneous steady state of the system; that
there is a local basis where the number of non-zero components of
the state vector coincides with the number of ``physical'' modes;
that in a system with finite number of degrees of freedom the
strict mode splitting disappears at finite value of coupling; that
above this value a fussy mode splitting is observed.
\end{abstract}

\maketitle

\section*{Introduction}
Nonlinear dissipative spatially extended systems have, from the
formal point of view, infinitely many degrees of freedom. But many
important examples are known where the chaotic solution of an
extended system evolves in an effective manifold of finite
dimension that is called the inertial manifold~\cite{Robinson}. H.
Yang et al.~\cite{EffDim} suggest that the tangent dynamics of
Kuramoto--Sivashinsky and Ginzburg--Landau equations is
essentially characterized by a well-defined set of vectors called
``physical'' modes which are decoupled from the remaining set of
hyperbolically ``isolated'' degrees of freedom. In this case the
physical modes can be a local linear approximation of the inertial
manifold, while isolated modes are orthogonal to this manifold and
are responsible only for transient processes.

The structure of the tangent space of a dynamical system is
characterized by Lyapunov exponents and associated with them
Lyapunov vectors. There are two orthogonal sets of vectors called
backward and forward Lyapunov vectors~\cite{AGuide}. They can be
computed in the course of the standard procedure of computation of
Lyapunov exponents~\cite{Benettin} in forward and backward time,
respectively~\cite{AGuide,ErshPotap}. These vectors are not
covariant with the tangent dynamics in a sense that the tangent
mapping, being applied to them, does not produce the forward or
backward vectors. Though the existence of the covariant Lyapunov
vectors (CLVs) was known for a long time, they became available
only recently thanks to effective numerical
algorithms~\cite{GinCLV,WolfCLV}. These vectors are not
orthogonal, they are invariant under time reversal and covariant
with the dynamics. Because these vectors span Oseledec subspaces,
they allow access to hyperbolicity properties~\cite{GinCLV}. CLVs
are the generalization of the notion of ``normal modes''. They are
reduced to the Floquet vectors if the flow is time periodic and to
the stationary normal modes if the flow is
stationary~\cite{WolfCLV}.

In this paper we study the mode splitting reported in
Ref.~\cite{EffDim}. The motivating idea is very simple. Consider
an extended dynamical system. When the spatial coupling is very
strong, the effective dynamics should be low-dimensional due to
synchronization effects. It means that the number of physical
modes should also be small. But when the coupling is very small,
the spatial cells become almost independent. In this case all
degrees of freedom are important so that the mode splitting can
vanish. We study a chain of amplitude equations that appear from
the Ginzburg--Landau equation when spatial discretization is
introduced. The step size of the discretization is used as a
control parameter. Varying the step we analyze the mode splitting
at different intensities of coupling.

The paper is organized as follows. In Sec.~\ref{sec:model} the
model system is described. Sec.~\ref{sec:iso_eig} is devoted to
the case of strong coupling when the system is close to the
bifurcation point. In Sec.~\ref{sec:strict} the strict mode
splitting is analyzed that is observed at moderate values of
coupling. Sec.~\ref{sec:fussy} represents the case of a weak
coupling when the strict splitting disappears and a fussy
splitting is observed instead. Finally, in Sec.~\ref{sec:sum} we
summarize the obtained results.

\section{The model system}\label{sec:model}

Consider a 1D complex Ginzburg--Landau equation $\partialt{a} = a
- (1+\myii c)|a|^2 a + (1+\myii b)\partialxx{a}$. To find
solutions of this equation numerically, we represent the second
spatial derivative as a finite difference. In this way, the
partial differential equation is transformed into a chain of $N$
amplitude equations:
\begin{equation}
\label{eqn:cgle} \dot{a}_n = a_n - (1+\myii c)|a_n|^2 a_n +
(1+\myii b)\kappa(a_n)/h^2,
\end{equation}
where $a_n\equiv a_n(t)$ ($n=0,1,\dots,N-1$) are complex
variables, $h$ is a step size of the discretization, and $c$ and
$b$ are real control parameters. We set $c=3$, $b=-2$ which
corresponds to the regime of so called ''amplitude
turbulence''~\cite{CgleChaos}. Function $\kappa(a_n)$ determines
the diffusive coupling and no-flux boundary conditions:
$\kappa(a_n) = a_{n-1}-2a_n+a_{n+1}$ $(n=1,2,\ldots N-2)$,
$\kappa(a_0) = 2(a_1-a_0)$, $\kappa(a_{N-1}) =
2(a_{N-2}-a_{N-1})$. We are interested in the properties of this
system at different strength of spatial coupling. So, the step
size $h$ shall be our control parameter. Treating the discrete
space representation of the Ginzburg--Landau equation as a chain
of oscillators allows us to freely change the step size $h$
without taking care of the validity of the numerical scheme.

To understand what happens when $h$ tends to zero, we perform the
rescaling in Eq.~\eqref{eqn:cgle} $h\to \epsilon h^\prime$, $a\to
a/\epsilon$, and $t\to \epsilon^2 t$. In the resulting equation
$\dot{a}_n = \epsilon^2 a_n - (1+\myii c)|a_n|^2 a_n + (1+\myii
b)\kappa(a_n)/(h^\prime)^2$ the decreasing of $\epsilon$
corresponds to the decreasing of $h$ in Eq.~\eqref{eqn:cgle}. The
$\epsilon$ here can be treated as a bifurcation parameter,
controlling the stability of the homogeneous steady
state~\cite{CgleChaos}. This state becomes unstable at
$\epsilon=0$, and the system enters the regime of spatio--temporal
chaos at $\epsilon>0$. So, returning to Eq.~\eqref{eqn:cgle}, we
can say that when $h$ is small the system is just a little bit
above the point of the emergence of spatio--temporal chaos, and it
has only a few positive Lyapunov exponent. Increasing $h$ results
in chaotic dynamics with an increasing number of positive Lyapunov
exponents.

\section{Isolated modes and eigenmodes}\label{sec:iso_eig}

Consider covariant Lyapunov vectors $\ell_i$ of the
system~\eqref{eqn:cgle}. When $h$ is decreased and the system
approaches from above the bifurcation point where the homogeneous
steady state becomes unstable, CLVs converge to eigenmodes of this
homogeneous steady state (in fact, these are the modes of Fourier
decomposition of the solution). For no-flux boundary conditions
the eigenmodes read $g_m(n)=s_m(t) \gamma_m(t) \cos(k_1 m n)$,
where $k_1=\pi/(N-1)$, and $m=0,\pm1, \dots \pm(N-2), N-1$. The
total number is $2N-2$, but because cosine is an even function,
modes $m$ and $-m$ are identical and only $N$ modes with $m\geq 0$
can be considered. $s_m(t)$ is a normalizing factor: $\sum_n
g_m(n)^2=1$. $\gamma_m(t)$ is a vector having two components which
can be computed using $\ell_i$. Vector $\ell_i$ has $N$ elements
$(\ell_i)_{2n}$ corresponding to $\Re(a)$ and also $N$ elements
$(\ell_i)_{2n+1}$ for $\Im(a)$. At the bifurcation point each
$\ell_i$ coincides with one of the eigenmodes, say $m$. It means
that dividing $(\ell_i)_{2n}$ and $(\ell_i)_{2n+1}$ by $\cos(k_1 m
n)$ we obtain a set of $N$ identical couples that are the
components of $\gamma_m(t)$. Beyond the bifurcation point, these
couples are not identical, and we define the vector $\gamma_m(t)$
as the average of them.

If the system is not far from the bifurcation point, $\ell_i$
should not diverge too much from $g_m$. To verify this, we compute
scalar products of each $\ell_i$ with each $g_m$ for many time
steps and find the average values. Both $\ell_i$ and $g_m$ are
normalized, hence, the scalar products are equal to cosines of the
angle between corresponding vectors. Two vectors of unit length
coincide when the cosine is equal to 1.

\begin{figure}
\onefig{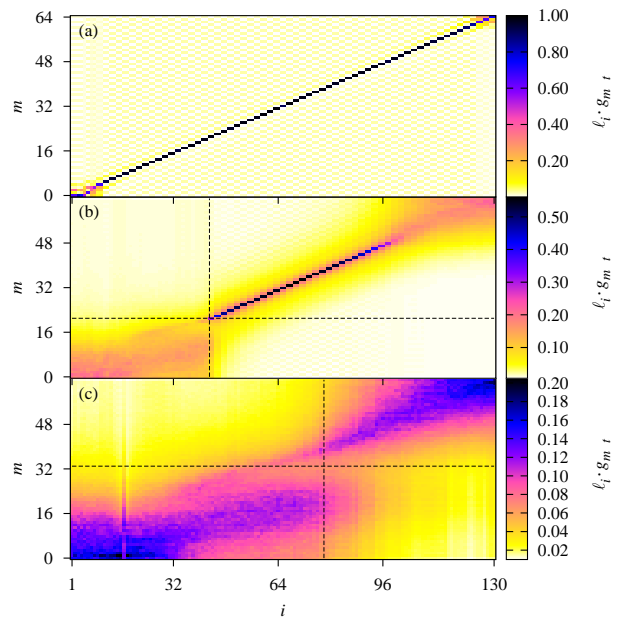}%
\caption{(color online) Average cosines of the angles between CLVs
$\ell_i$ (horizontal axis) and eigenmodes $g_m$ (vertical axis).
(a) $h=0.1$; (b) $h=0.5$, $i=42$ for the vertical dashed line and
$m=21$ for the horizontal one; (c) $h=0.8$, $i=77$, $m=33$.}
\label{fig:cgle_eig}
\end{figure}

Figure~\ref{fig:cgle_eig}(a) show average cosines at $h=0.1$ when
the system is close to the bifurcation point and has only one
positive Lyapunov exponent $\lambda_1\approx 0.084$. A large part
of the diagram is occupied by the black points along the diagonal
surrounded by white area. (In fact, there are pale squares off the
diagonal, but this is a numerical artifact.) It means that
corresponding $\ell_i$ indeed coincide with $g_m$. Notice that the
points are grouped pairwise. This is a manifestation of the above
mentioned degeneracy of eigenmodes with $m$ and $-m$. The
non-degenerated modes are orthogonal to each other and are
referred to as isolated modes~\cite{EffDim}.

The degeneracy of isolated modes is associated with the degeneracy
of eigen modes that, in turn, depends on the geometry of the
system. In our case the no-flux boundary conditions leads for the
Ginzburg--Landau equation to a degeneracy of the order two, while
in~\cite{EffDim} periodic boundary conditions give rise to a
four-fold degeneracy.

There is an area in the left bottom corner of
Fig.~\ref{fig:cgle_eig}(a) where CLVs differ significantly from
the eigenmodes. In Ref.~\cite{EffDim} such kind of vectors has
been called ``physical''. We shall refer to them as active
vectors. These vectors are discussed in the following section.

\section{Strict mode splitting}\label{sec:strict}

\subsection{Angles with eigenmodes}

The number of active vectors at $h=0.1$ is small because the
system is close to the bifurcation point. In
Fig.~\ref{fig:cgle_eig}(b) $h=0.5$, and the system has 9 positive
Lyapunov exponents. We observe now a large area of active vectors
that is clearly separated from the set of isolated vectors. The
isolated vectors are represented by the diagonal structure. The
diagonal is not so sharp as in panel (a), which means that now
angles between isolated vectors and corresponding eigenmodes,
though small, are not equal to zero. Correspondingly, these
vectors are not quite orthogonal to all other eigenmodes. But
nevertheless, the isolated vectors remain very close to the eigen
modes. The split between isolated and active modes is marked by
the vertical dashed line at $i=42$. Also, the area of active modes
is bounded from above: the horizontal dashed line is drawn at
$m=21$. It means that the active vectors have relatively small
angles only with ``their own'' eigenmodes, i.e., with eigenmodes
with numbers corresponding to the active vectors. The angles with
the other eigenmodes are much higher. Thus, the set of active
vectors span approximately the same subspace as the corresponding
amount of the eigenmodes.

There is another non-trivial structure at the right top corner of
Fig.~\ref{fig:cgle_eig}(b). The nature of this area is unclear
yet, but we conjecture that it consists of active vectors that
becomes relevant when time is reversed.

\subsection{Fraction of DOS violation}

The isolated modes do not have tangencies with the active
modes~\cite{EffDim}. The method of detection of this strict mode
splitting, suggested in Ref.~\cite{EffDim}, employs a concept of
domination of Oseledec splitting (DOS)~\cite{DOS1,DOS2}. We recall
that for almost every time every vector in the tangent space
$\oslspace_1(t)$ of a dynamical system grows asymptotically at
rate given by the first Lyapunov exponent $\lambda_1$ except those
belonging to a set $\oslspace_2(t)$ of measure zero. Similarly,
almost every vector in $\oslspace_2(t)$ asymptotically grows at
rate $\lambda_2$ except those belonging to a set $\oslspace_3(t)$
of measure zero relative to $\oslspace_2(t)$, and so on.
Collection of sets $\oslspace_i(t)$ embedded one into another is
called the Oseledec splitting of the tangent space. The splitting
is called dominated if each Oseledec subspace is more expanded
than the next, by a definite uniform factor. Let
$\lambda_i(t,\tau)$ be the $i$-th local Lyapunov exponent,
computed at time $t$ and averaged over an interval $\tau$. The
Oseledec splitting is dominated at $i$ if
$\lambda_{i}(t,\tau)>\lambda_{j}(t,\tau)$ holds for all $j>i$, and
for all $t$ with $\tau$ larger then some finite
$\tau_0$~\cite{EffDim}. In particular, domination implies that the
angles between the Oseledec subspaces are bounded from
zero~\cite{DOS2}.

Employing the ideas of numerical verification of DOS suggested in
Refs.~\cite{Psplit,DOSV,EffDim,YRRev} we define the fraction of
DOS violation in the following manner. Fix an interval $\tau$ and
compute $\lambda_i(t,\tau)$ for some time $t$. The violation of
DOS takes place if $\lambda_i(t,\tau)\leq \lambda_j(t,\tau)$ for
$j>i$. Thus, for each $i$ we check this inequality at $j>i$ and
add 1 to the $i$-th site of an array if it holds at least ones.
Repeating this procedure for different times and performing an
averaging we obtain the fraction of DOS violation at $\tau$, which
formally can be defined as
$\fdosv=\left<\Theta\left(\max_{j>i}\left[\lambda_j(t,\tau)-
\lambda_i(t,\tau)\right]\right)\right>_t$, where $\Theta(z)$ is
the step function and $<\ldots>_t$ denotes the time average.

In the case of multiplicity $\lambda_i=\lambda_{i+1}$,
corresponding $\fdosv$ is close to 0.5 because of fluctuations of
local Lyapunov exponents due to a numerical noise. Otherwise
$\fdosv$ decays to zero as $\tau$ grows. In general, the decay is
asymptotic, but if the splitting is dominated at $i$, the
corresponding fraction $\fdosv$ vanishes at finite $\tau_0$.
Unfortunately, there is no a well-grounded algorithm of
computation of $\tau_0$ except the straightforward observation of
$\fdosv$ as a function of $\tau$. Because the law of the decay of
$\fdosv$ is unknown a priori, there is no idea how to extrapolate
of $\fdosv$ to zero to verify if a finite $\tau_0$ exists. But,
anyway, in points of dominated splitting $\fdosv$ decays much
faster against $\tau$ than elsewhere. It means that if $\tau$ is
sufficiently large, a graph $\fdosv$ against $i$ provides relevant
information about locations of the splitting.

\begin{figure}
\onefig{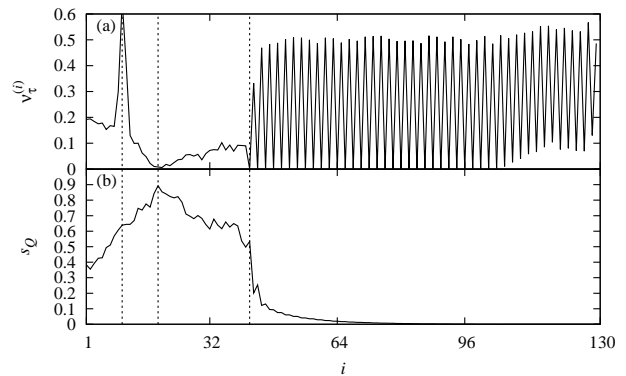}%
\caption{(a) Fraction of DOS violation at $\tau=51.2$ vs. number
of Lyapunov exponent. (b) Average projection of the state vector
onto vector-columns of $Q$ vs. the number of vector. Dotted
vertical lines marks, from left to right: index of the first of
two zero Lyapunov exponents $i=10$, Kaplan--Yorke (Lyapunov)
dimension rounded up to the next integer $i=19$, and the number of
active vectors $i=42$.}%
\label{fig:cgle_dp_05}
\end{figure}

\begin{figure}
\onefig{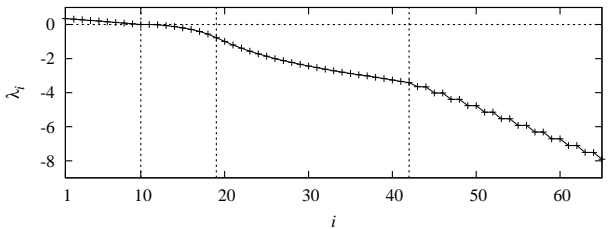}%
\caption{Lyapunov spectrum at $h=0.5$. The horizontal dashed line
mark $\lambda_i=0$, and the vertical ones are drawn at the same
positions as in
Fig.~\ref{fig:cgle_dp_05}}%
\label{fig:cgle_lyap}
\end{figure}

Figure~\ref{fig:cgle_dp_05}(a) shows the fraction of DOS violation
at $h=0.5$. The sharp minimum of $\fdosv$ at $i=42$ coincides with
the position of the splitting found in Fig.~\ref{fig:cgle_eig}(b).
This minimum, presuming the vanish of $\fdosv$ at a finite $\tau$,
means that the active modes are hyperbolically isolated from all
the rest ones. The active vectors, located to the left from the
splitting point, have sufficiently high $\fdosv$. In this case
$\fdosv$ vanishes only asymptotically which, in turn, indicates
frequent tangencies between the active vectors. The isolated
vectors are represented by a series of sharp minima and maxima
with the period 2. Above we have shown that these vectors at
$h=0.5$ are very close to the eigenmodes $g_m$. Because of the
degeneracy, the modes $g_m$ and $g_{-m}$ have identical growth
rates. Hence, there are couples of corresponding isolated vectors
$\ell_i$ with identical growth rates. In turn, this implies the
multiplicity of corresponding Lyapunov exponents. Indeed, the
spectrum of Lyapunov exponents demonstrates a stepwise behavior to
the right from the splitting point $i=42$ and the step is 2, see
Fig.~\ref{fig:cgle_lyap}. (The stepwise structure of the Lyapunov
spectrum corresponding to isolated vectors was also reported in
Ref.~\cite{EffDim}.) Thus, the maxima of $\fdosv$ in
Fig.~\ref{fig:cgle_dp_05}(a) are associated with this
multiplicity. The deep minima indicate the absence of tangencies
between isolated vectors because of the orthogonality of
corresponding eigenmodes.

The curve of $\fdosv$ in Fig.~\ref{fig:cgle_dp_05}(a) demonstrates
two more interesting features. We observe another point of
splitting at $i=19$ where $\fdosv$ has a very deep minimum. The
Kaplan--Yorke (Lyapunov) dimension in this case is
$\lyapdim\approx 18.2$. Thus, we can conjecture, that there are
two types of active modes that are hyperbolically isolated from
each other, and the amount of the first type modes is equal to the
Kaplan--Yorke dimension rounded up to the next integer.

Also notice the sharp spike at $i=10$. To explain the emergence of
this spike we need to recall that the complex Ginzburg--Landau
equation with no-flux boundary conditions has two zero Lyapunov
exponents~\cite{SymCGLE1,*SymCGLE2}. There are 9 positive
exponents and $\lambda_{10}=\lambda_{11}=0$, so the spike at
$i=10$ indicates the multiplicity of two corresponding~$\ell_i$.

\subsection{Projections of the state vectors}

The split of CLVs onto active and isolated vectors should have a
clear and visible manifestation in the dynamics of a system. The
idea that the number of active vectors $i$ is an effective
dimension of the system, suggested in~\cite{EffDim}, presumes that
there exists a local basis where the state vector has only $i$
nonzero components. We consider projection of the state vector of
the system at time $t$ on the local basis composed of backward
Lyapunov vectors. These vectors are orthogonal to each other and
span the same subspaces as CLVs~\cite{AGuide}. The backward
Lyapunov vectors can be computed much faster then CLVs in the
course of the standard procedure of computation of the Lyapunov
exponents~\cite{Benettin}, because the columns of orthogonal
matrices $Q$ converge to them~\cite{AGuide,ErshPotap}. So, we
multiply transposed matrices $Q^\text{T}$ by corresponding state
vectors, accumulate absolute values of obtained projections, and
then average them over large number of steps. (In general case,
the homogeneous steady state should be subtracted from the state
vector before the multiplication to avoid an unnecessary shift.
But for our system the homogeneous steady state is 0.)

The average projection denoted as $s_Q$ is shown in
Fig.~\ref{fig:cgle_dp_05}(b). We observe the curve that agrees
good with $\fdosv$. Projections onto active vectors are large,
while they are almost zero for the isolated vectors. These two
parts of the curve are clearly separated exactly at $i=42$. This
clarifies the nature of the discussed mode splitting: We indeed
observe that the number of active vectors can be an effective
dimension of the system. Also notice that the largest component of
the projection has an index which is equal to the Kaplan--Yorke
dimension rounded up to the next integer, as marked by the middle
dashed line in the figure.

The idea of using backward Lyapunov vectors in dimension reduction
methods has already been considered and rejected as non-promising
in Refs.~\cite{Proj1,Proj2,Proj3}. Indeed, the validity of this
basis is not so obvious. Reasoning formally, one can imagine an
attractor that has inappropriate orientation in the phase space so
that the decomposition fails to give correct result. But on the
other hand, let us assume that we have a small spherical cloud of
points that surrounds a homogeneous steady state in the phase
space. When the points evolve, the cloud is extended along the
most unstable manifold. In this case the most information about
dynamics is positively carried by the first several CLVs. We can
guess, that this property survives later producing the split into
active and isolated vectors. Anyway, at least for the
Ginzburg--Landau equation this decomposition gives very
appropriate information concerning the mode splitting.

The picture illustrated in this section for $h=0.5$ is quite
generic. We can observe the strict splitting of active and
isolated modes as well as the Kaplan--Yorke mode splitting for a
wide range of $h$. But when $h$ becomes sufficiently high, the
situation becomes quite different, as will be discussed in the
following section.

\section{Fussy mode splitting}\label{sec:fussy}

\begin{figure}
\onefig{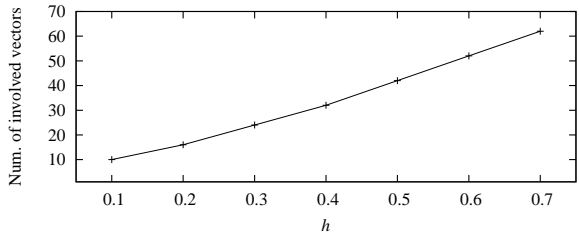}%
\caption{Number of active vectors against the coupling parameter
$h$.}%
\label{fig:hspl}
\end{figure}

As discussed above, exactly at the bifurcation point CLVs coincide
with eigenmodes and all of them are isolated, while the set of
active vectors is empty. There are $2N-2$ isolated modes. When $h$
grows, the isolated modes are converted into active ones and this
conversion occurs at both ends of the spectrum. (Compare small
structures at the ends of the main diagonal in
Fig.~\ref{fig:cgle_eig}(a) and the large areas in
Fig.~\ref{fig:cgle_eig}(b).) At the left end the isolated modes
contribute to the set of active modes, while at the right end they
fill up the other set of modes, which are, conceivably, relevant
when time is reversed. Figure~\ref{fig:hspl} shows that the number
of active modes depends on $h$ almost linearly. Because we have a
finite number of modes, there is a finite $h$ for which all
isolated modes are converted so that the splitting vanishes. If
the conversion at both ends of the spectrum takes place
symmetrically, then the isolated modes disappear when there are
$N-1$ active modes. In Fig.~\ref{fig:hspl} the splitting of modes
indeed disappears when the number of active modes is close to 64
at $N=65$.

The value of $h$ where the mode splitting disappears depends on
the number of eigenmodes, that, in turn depends on the number of
oscillators in the chain $N$. Taking into account almost linear
dependance of the number of active modes against $h$ we conclude
that in a chain with an infinite number of oscillators the mode
splitting vanishes at infinite $h$. Thus, the continuous system
can have the mode splitting at any strength of the spatial
coupling. In particular, it gives a criterion of correctness of a
chain as a model of continuous system: the chain can model a
continuous system only if the step size is below the point where
the mode splitting vanishes.

Now we consider the tangent space above the point of mode
splitting vanishing. Figure~\ref{fig:cgle_eig}(c) demonstrates
angles between CLVs and eigenmodes at $h=0.8$ (there are 17
positive Lyapunov exponents in this case). One can see that this
figure differs much from the panels (a) and (b). Sets of vectors
are still distinguishable, but their boundaries are not strict.
The isolated vectors are absent at all. The vertical dashed line
marks an approximate boundary between two clusters of vectors, and
the horizontal one separate the spectrum of eigenmodes onto two
halves. The vectors from the left cluster have relatively small
angles with the first half of the spectrum of eigenmodes, while
the right cluster contains the vectors that have relatively small
angles with the second half of the spectrum. In particular, it
means that tangencies within the clusters occur more often than
between the representatives of two clusters. It can be treated as
a fussy mode splitting: The left cluster is preliminary relevant
to the forward time dynamics, while the right one, though also
involved, but does not include much. In the reverse time the roles
of the clusters are exchanged.

\begin{figure}
\onefig{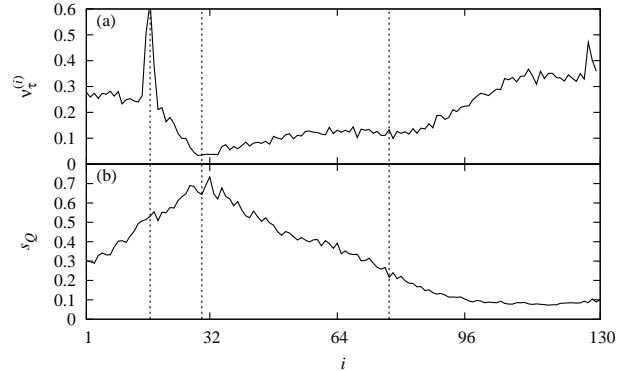}%
\caption{Same as Fig.~\ref{fig:cgle_dp_05} at  $h=0.8$. The
vertical dashed lines are plotted at $i=17$, $i=30$ and $i=77$.}%
\label{fig:cgle_dp_08}
\end{figure}

Figure~\ref{fig:cgle_dp_08} shows the fraction of DOS violation
and average projections of the state vector onto backward Lyapunov
vectors at $h=0.8$. There is no series of sharp minima and maxima
representing isolated modes, and also there is no sharp step in
the curve of projections. But the curve of projections indicates
that the state vector still has preferable directions inside the
left cluster of CLVs, whose boundary is marked by the vertical
dashed line. In the other words, the number of vectors in the left
cluster can be considered as approximate effective dimension of
the system. The projections to vectors from the right cluster is
much smaller.

Notice also that two other features of these curves survive. We
still can see the minimum of $\fdosv$ and the maximum of the
projection corresponding to the Kaplan--Yorke dimension rounded up
to the next integer, as well as the spike, indicating degeneracy
associated with two zero Lyapunov exponents.

\section{Summary}\label{sec:sum}

We studied the splitting of modes of perturbations, represented by
the covariant Lyapunov vectors, into sets of active and isolated
modes~\cite{EffDim}. We considered a chain of amplitude equations
obtained from the Ginzburg--Landau equation by substitution of the
second spatial derivative with its finite--difference
representation. The size of the step of spatial discretization was
used as a control parameter while the number of oscillators in the
chain was held constant. When the step size is asymptotically
small, the system approaches from above the bifurcation point
where the homogeneous steady state looses its stability, while
increasing of the step results in more independent dynamics of
oscillators.

At the bifurcation point there are no active modes. All modes are
isolated and coincide with eigenmodes of the homogeneous steady
state. Their spatial structure is determined by the number of
oscillators and boundary conditions. When the system leaves the
bifurcation point as the step size grows, the isolated modes are
converted into active ones so that the number of active modes
grows linearly with the step size.

For the considered system, the backward Lyapunov vectors was shown
to be an appropriate basis where a number of essential components
of the state vector is equal to the number of active vectors. In
other words, the number of active vectors indeed plays the role of
an effective dimension of the system, as conjectured
in~\cite{EffDim}.

The active modes were found to be split into two subsets that are
hyperbolically isolated from each other. The coordinate of the
splitting point is equal to the Kaplan--Yorke dimension rounded up
to the next integer. We conjecture that this indicates the
existence of two types of active modes. The nature of these
different types is unclear yet, and more studies are required.

At the right end of the spectrum we observed another set of modes
which is similar to the set of active modes. Its nature is unclear
yet, but we conjecture that these modes become relevant when time
is reversed.

At a certain finite value of the step size the strict mode
splitting disappears. Because this value depends on the number of
oscillators in the chain, the vanish of the splitting occurs only
for a system with finite number of degrees of freedom and probably
can not be observed, in particular, for continuous systems. It can
be used as an estimation of the maximum step size of the spatial
discretization. If a continuous system has the strict mode
splitting and its discrete model does not have it, it means that
the step size is too large.

Above the point where the splitting vanishes the spectrum of modes
contains two clusters without strict boundaries. This can be
treated as a fussy mode splitting. The first cluster contains
formerly active modes, while the other one corresponds,
apparently, to the modes mainly involved when time is reversed.
The projection of the state vector on the backward Lyapunov
vectors indicates that the number of vectors in the first cluster
could be an approximation of an effective dimension of the system.

\begin{acknowledgments}
Authors acknowledge valuable discussions with A. Pikovsky, A.
Politi and P. Cvitanovi\'{c}. PVK acknowledges support from
Deutscher Akademischer Austausch Dienst and Russian Ministry of
Education and Science, program ``Michail Lomonosov II''. UP thanks
the Max--Plank--Society for financial support.
\end{acknowledgments}

\bibliography{ts}

\end{document}